\documentclass[%
 floatfix,
 reprint,
 twocolumn,
superscriptaddress,
showpacs,preprintnumbers,
 amsmath,amssymb,
 aps,
prd,
]{revtex4-2}

\usepackage[colorlinks = true,
            linkcolor = red,
            urlcolor  = blue,
            citecolor = red,
            anchorcolor = blue]{hyperref}

\usepackage[caption=false]{subfig} 
\usepackage{makecell}

\usepackage{graphicx}  
\usepackage{dcolumn}   
\usepackage{bm}        
\usepackage{amssymb}   
\usepackage{comment} 
\usepackage{color}
\usepackage[percent]{overpic}
\usepackage{amsmath}
\usepackage{footmisc}
\usepackage{booktabs}
\usepackage{xspace}
\usepackage{gensymb}
\usepackage[normalem]{ulem}

\newcommand{\nnbar}{$n$-$\bar{n}$\xspace}

\begin{document}

\widetext
\title{First Observation of Antiproton Annihilation At Rest on Argon in the LArIAT Experiment}

\newcommand{\UFederalABC}{Universidade Federal do ABC, Santo Andr\'{e}, SP 09210-580, Brasil}
\newcommand{\UCampinas}{Universidade Estadual de Campinas, Campinas, SP 13083-859, Brasil}
\newcommand{\UChicago}{University of Chicago, Chicago, IL 60637, USA}
\newcommand{\cincinnati}{University of Cincinnati, Cincinnati, OH 45221, USA}
\newcommand{\FNAL}{Fermi National Accelerator Laboratory, Batavia, IL 60510, USA}
\newcommand{\UFG}{Universidade Federal de Goi\'{a}s, Goi\'{a}s, CEP 74690-900, Brasil}
\newcommand{\UTArlington}{University of Texas at Arlington, Arlington, TX 76019, USA} 
\newcommand{\Boston}{Boston University, Boston, MA 02215, USA}
\newcommand{\michigan}{Michigan State University, East Lansing, MI 48824, USA}
\newcommand{\duluth}{University of Minnesota, Duluth, Duluth, MN 55812, USA}
\newcommand{\infn}{Istituto Nazionale di Fisica Nucleare (INFN), Rome 00186, Italy}
\newcommand{\louisiana}{Louisiana State University, Baton Rouge, LA 70803, USA}
\newcommand{\Syracuse}{Syracuse University, Syracuse, NY 13244, USA}
\newcommand{\utaustin}{University of Texas at Austin, Austin, TX 78712, USA}
\newcommand{\ucollegelondon}{University College London, London WC1E 6BT, UK}
\newcommand{\williamandmary}{College of William \& Mary, Williamsburg, VA 23187, USA}
\newcommand{\Yale}{Yale University, New Haven, CT 06520, USA}
\newcommand{\kek}{High Energy Accelerator Research Organization (KEK), Tsukuba 305-0801, Japan}
\newcommand{\IIT}{Illinois Institute of Technology, Chicago, IL 60616, USA}
\newcommand{\LANL}{Los Alamos National Laboratory, Los Alamos, NM 87545, USA}
\newcommand{\Manchester}{University of Manchester, Manchester M13 9PL, UK}
\newcommand{\Harvard}{Harvard University, Cambridge, MA 02138, USA}
\newcommand{\UCSB}{University of California, Santa Barbara, CA 93106, USA}
\newcommand{\UCDavis}{University of California, Davis, CA 95616, USA}
\newcommand{\Edinburgh}{University of Edinburgh, Edinburgh, EH8 9YL, UK}
\newcommand{\krakow}{Institute of Nuclear Physics PAN, 31-342 Krak\'{o}w, Poland}
\newcommand{\dallas}{University of Dallas, Irving, Texas 75062, USA}


\author{V.~Basque} \affiliation{\FNAL}
\author{R.~Acciarri} \affiliation{\FNAL}
\author{J.~Asaadi}\affiliation{\UTArlington}
\author{M.~Backfish}\altaffiliation{Current:~\UCDavis}\affiliation{\FNAL}
\author{W.~Badgett}\affiliation{\FNAL}
\author{F.~Cavanna}\affiliation{\FNAL}
\author{W.~Flanagan}\affiliation{\dallas}
\author{W.~Foreman} \altaffiliation{Current:~\LANL} \affiliation{\IIT}
\author{R.~A.~Gomes} \affiliation{\UFG}
\author{E.~Gramellini}\affiliation{\Manchester}
\author{M.~A.~Hernandez-Morquecho}\affiliation{\IIT}
\author{J.~Ho} \altaffiliation{Current:~\Harvard}\affiliation{\UChicago}
\author{E.~Kearns} \affiliation{\Boston}
\author{E.~Kemp} \affiliation{\UCampinas}
\author{M.~King} \affiliation{\UChicago}
\author{T.~Kobilarcik}\affiliation{\FNAL}
\author{P.~Kryczy\'nski}\affiliation{\krakow}
\author{B.~R.~Littlejohn} \affiliation{\IIT}
\author{A.~Marchionni} \affiliation{\FNAL}
\author{C.~A.~Moura} \affiliation{\UFederalABC}
\author{J.~L.~Raaf} \affiliation{\FNAL}
\author{D.~Schmitz} \affiliation{\UChicago}
\author{M.~Soderberg}\affiliation{\Syracuse}
\author{J.~M.~St.~John}\affiliation{\FNAL}
\author{A.~M.~Szelc}\affiliation{\Edinburgh}

\collaboration{LArIAT Collaboration}
\thanks{lariat\_authors@fnal.gov}\noaffiliation       

\date{\today}

\begin{abstract}

We report the first observation and measurement of antiproton annihilation at rest on argon track and shower multiplicities and particle identification conducted with the LArIAT experiment. Stopping antiprotons from the Fermilab Test Beam Facility's charged particle testbeam are identified using beamline instrumentation and LArIAT’s liquid argon time projection chamber (LArTPC). The charged particle multiplicity from the annihilation vertex is manually evaluated via hand-scanning, yielding a mean of 3.2~$\pm$~0.4~tracks and a standard deviation of 1.3~tracks, consistent with a semi-automated reconstruction resulting in 2.8~$\pm$~0.4~tracks and a standard deviation of 1.2~tracks. Both methods are consistent with Monte Carlo simulations within statistical uncertainty. The shower multiplicities and particle identification for outgoing tracks are also consistent with Geant4 models predictions. These results, obtained from a low-statistics sample, provide a foundation for higher-statistics studies in larger LArTPCs, which could refine modeling of intranuclear annihilation on argon and inform scenarios such as neutron-antineutron oscillations.

\end{abstract}

\maketitle

\section{Introduction}

Due to their distinct ``star-shaped'' final-state topology, antiproton ($\bar{p}$) annihilations at rest have produced some of the most striking images in particle physics. First observed in 1955 at the Bevatron using photographic emulsion~\cite{PhysRev.100.947, longer_chamberlain, PhysRev.105.1037}, $\bar{p}$ annihilation on nucleons ($N$) has been observed by a number of experiments, and on various target nuclei~\cite{DIAZ1970239, PhysRev.139.B1659, PhysRev.140.B1039, PhysRev.140.B1042, Chiba:1987ge, Chiba:1989yw, AHMAD1985135, Aker:1992ny, AGNELLO199711, Amsler:1997up, CrystalBarrel:2003uej, Aghion_2014, Aghion_2017}.

The $\bar{p}N$ annihilation takes place preferentially at the surface of the nucleus~\cite{VonEgidy:1987mz} and results in a set of particles, primarily pions ($\pi$), emanating isotropically from the annihilation vertex. The exact number and mixture of produced particle types varies depending on the target nucleus and its density. Since a fraction~\cite{VonEgidy:1987mz, antiproton_confinment, PhysRevC.63.027301} of the pions from the annihilation have to travel through the nucleus before becoming visible in the detector, they can undergo a number of final-state interactions (FSI) processes. These processes modify the kinematics of the pions or can prevent pions from exiting the nucleus entirely.

Assuming isospin symmetry, and the similarity of protons and neutrons in terms of quark content, it can be assumed that $\bar{p}N$ annihilation interactions result in outgoing products that are largely similar in particle makeup and kinematic distribution to those that would emerge from a $\bar{n}N$ annihilation~\cite{Klempt:2005pp}. Because of this, $\bar{p}N$ annihilation can be used as a proxy for that of theorized neutron-antineutron (\nnbar) oscillations~\cite{Mohapatra:2009wp} where the newly-converted $\bar{n}$ immediately annihilates with a neighboring nucleon. Observing this baryon number violating process would satisfy one of the Sakharov conditions~\cite{Sakharov:1967dj} and would have profound implications for theories on baryogenesis, the origin of the matter-antimatter asymmetry, and large extra dimensions~\cite{FUKUGITA198645, PhysRevLett.97.131301, PhysRevD.87.115019, PhysRevLett.88.171601}. Due to the difficulty in performing antineutron scattering measurements and their resulting scarcity~\cite{Astrua:2002zg}, $\bar{p}N$ annihilation are critical inputs for \nnbar oscillation models by providing more final-state information. More complete models, especially of FSI due to the nucleus, are necessary improve the sensitivity of future searches for \nnbar in large liquid argon time projection chambers (LArTPCs) like the Deep Underground Neutrino Experiment (DUNE)~\cite{DUNE:2020lwj}.

This paper presents the first measurement of antiproton annihilation at rest on argon, performed with the LArIAT liquid argon testbeam experiment~\cite{Acciarri:2019wgd}. Antiprotons are identified amongst the particles produced in the beamline. A subset of the antiprotons are identified as annihilating at rest and the outgoing particles emerging from the annihilation are compared to a Monte Carlo (MC) simulation. The results presented here represent a first validation of the modeling of nucleon annihilation on argon and may be useful for constraining systematic uncertainties of FSI for pions produced by nucleon annihilation on argon.

\section{The LArIAT Experiment}\label{sec:lariat_detector}

The LArIAT experiment collected data at Fermilab's Test Beam Facility~\cite{testbeam_fac} situated in the Meson Center beam line. LArIAT's primary purpose is measuring charged particle interaction cross sections and developing calibration techniques. The LArIAT TPC operated in conjunction with a series of beamline monitoring subsystems to enable the triggering of detector readouts for individual particles that traverse the entire beamline and interact within the chamber. 

\begin{figure*}[!ht]
\includegraphics[width=0.9\textwidth,height=\textheight,keepaspectratio]{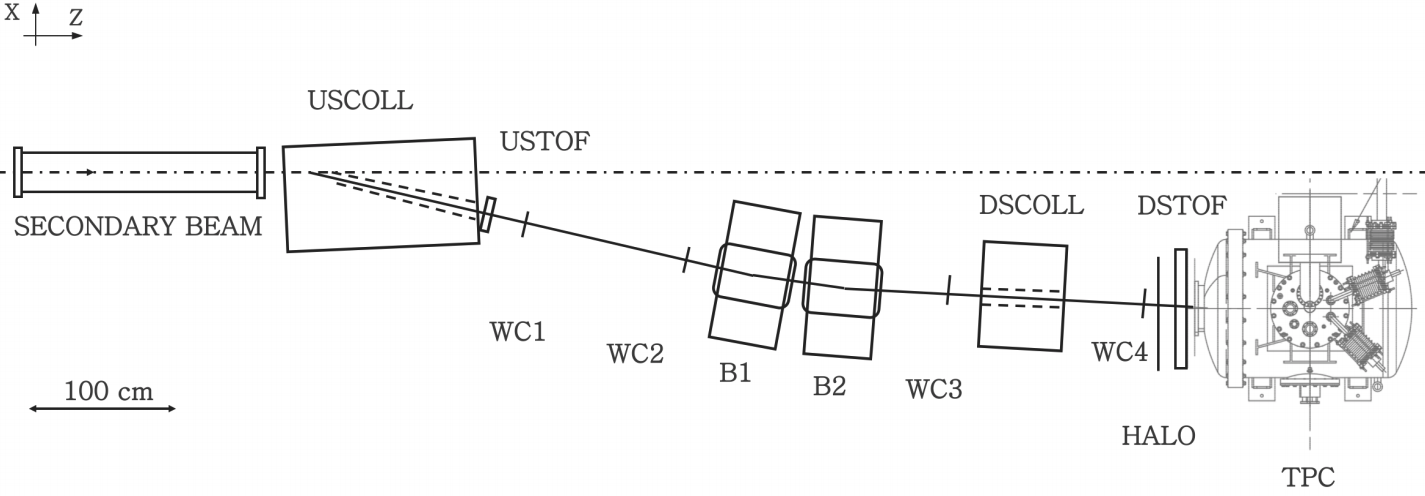}
\caption{%
  Top-down view of LArIAT's tertiary beamline which shows the location and orientation of the beamline monitors and a technical drawing of the LArTPC. The monitors are labeled as follows: upstream and downstream beam collimators (USCOLL and DSCOLL), upstream and downstream time-of-flight scintillating paddles (USTOF and DSTOF), four multi-wire proportional chambers (WC1-WC4), two bending dipole magnets (B1 and B2), and a HALO scintillating paddle sensitive to particles which did not follow the path of the tertiary beamline. Figure is from~\cite{Acciarri:2019wgd}.
}\label{fig:beamline_cartoon}
\end{figure*}

Fermilab's accelerator complex produces a 120~GeV/c primary proton beam which is focused onto a copper target to produce a secondary beam composed of primarily positively charged pions. The secondary beam momentum can be tuned between the range of 8~GeV/c to 80~GeV/c and was operated at 64~GeV/c for the datasets used in this analysis.

Before reaching LArIAT, the secondary beam entered the experimental hall enclosure where it struck a copper target inside a steel collimator to produce a tertiary beam of charged particles. The tertiary beam traveled through a series of beamline instrumentation elements as shown in Fig.~\ref{fig:beamline_cartoon}, which included two time-of-flight (TOF) scintillating paddles and four multi-wire proportional chambers referred to as `wire chambers' (WCs). The trajectory of a charged particle traveling in the beamline was reconstructed by interpolating the straight segments obtained by at least three hits in the upstream WCs in tandem with the downstream ones. This information was used to reconstruct each particle’s three-dimensional trajectory along the tertiary beamline. The polarity of the dipole magnets could be configured to select positively or negatively charged particles. The bending angle and known magnetic field were used to determine the particle’s momentum. In addition, the scintillating paddles measured the particle’s TOF, which was used in tandem with the momentum to determine the particle mass as explained in Sec. \ref{sec:beamline}. Detailed discussion on the TOF and WC systems can be found in~\cite{Acciarri:2019wgd, LArIAT:2021yix}.

\begin{figure}[tb]
	\centering
	\includegraphics[width=0.80\columnwidth]{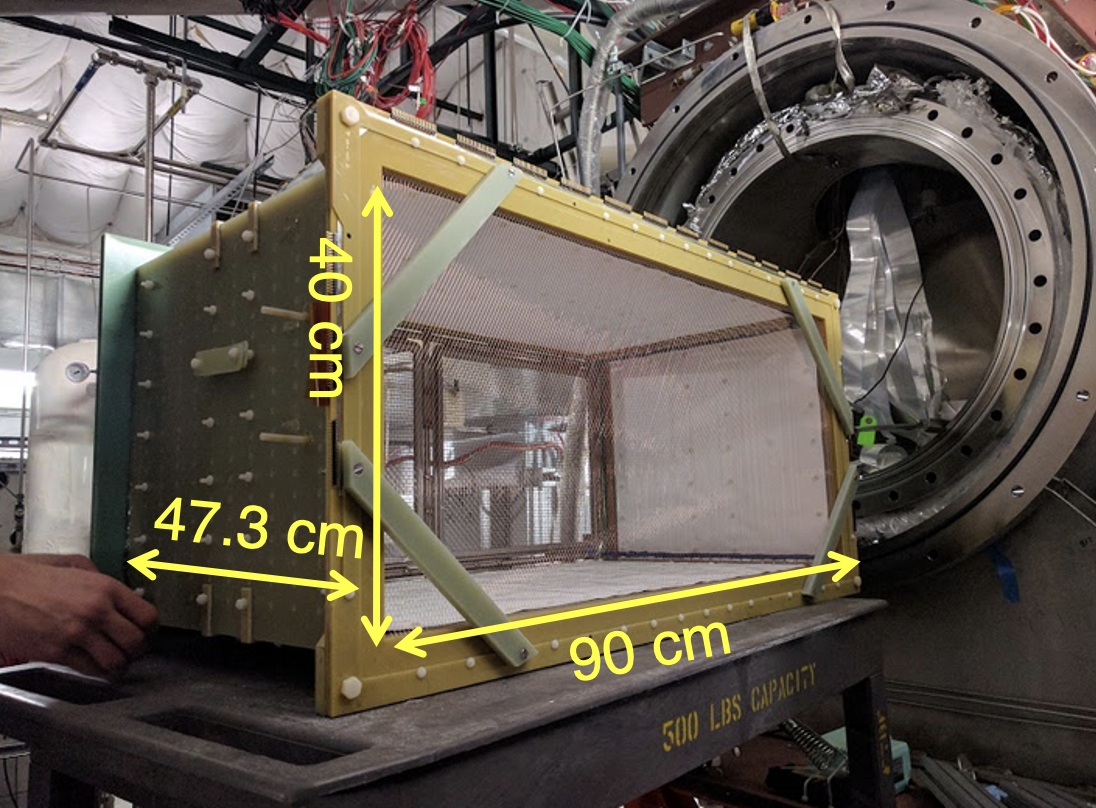}
	\caption{Photograph of the LArIAT TPC including its dimensions sitting in front of the cryostat.}
	\label{fig:lariat_tpc_photos}
\end{figure}

Tertiary beamline particles were then directed into the LArTPC, a 170-liter active LAr volume approximately 90~cm in length ($\hat{z}$, along the beam direction), 47~cm in width ($\hat{x}$, extending from the anode to the cathode), and 40~cm in height ($\hat{y}$, pointing upward). The LArTPC and cryostat are shown in Fig.~\ref{fig:lariat_tpc_photos}. The LArTPC readout comprised two readout-instrumented wire planes composed of 240 wires each (induction and collection), as well as a non-instrumented 225-wire shield plane. For the data used in this measurement, wires were spaced 4~mm apart with a 4~mm separation gap between each of the planes. The induction and collection wire planes were oriented at $\pm$60\degree{} relative to the beamline direction. Ionization charge drifted along the $\hat{x}$ direction from the cathode to the anode wire planes in a nominal field of 484~V/cm, which resulted in a maximum electron drift period of approximately 320~$\mu$s. The LArIAT data acquisition system recorded the ionization produced by beamline particles traversing the LArTPC via induced or collected charge on the wire planes, and these signals were digitized with a sampling period of 128~ns.

The analysis presented here uses two weeks of beam-triggered data, taken during LArIAT’s `Run~II’ data-taking period in 2016, when the tertiary beamline’s dipole magnets were operating in negative-polarity mode in order to steer negatively-charged particles toward the TPC.

\section{Beamline Reconstruction and Antiproton Selection}\label{sec:beamline}

\begin{figure}[tb]
	\centering
	\includegraphics[width=0.93\columnwidth, trim=0cm 0cm 0cm 1.5cm, clip=true]{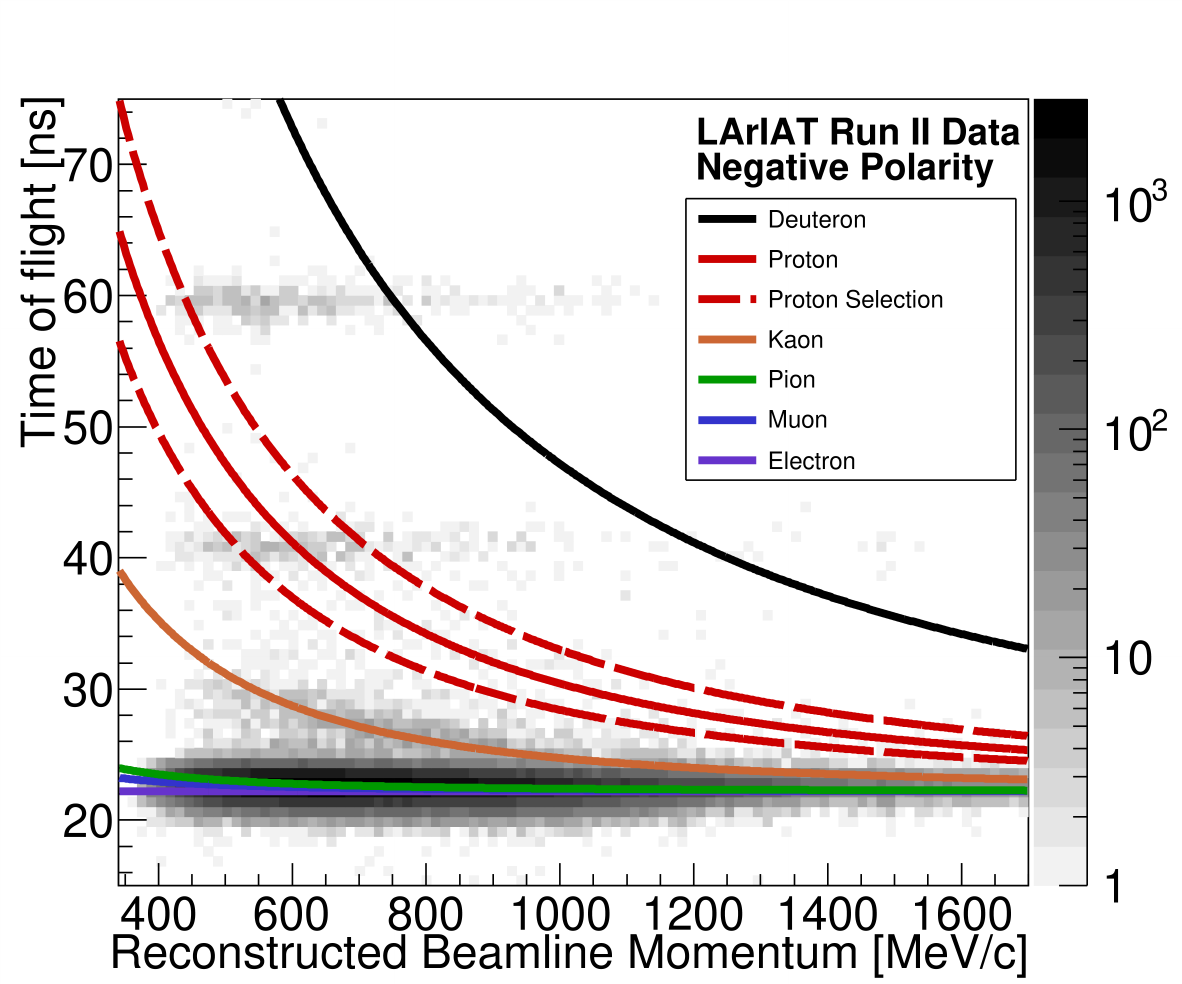}
	\caption{Reconstructed beamline momentum and TOF for Run~II negative-polarity data overlaid with the expectation curves of electrons in purple, muon in blue, pion in green, kaons in orange, protons in red, and deuterons in black. The dashed lines show from the mass selection cut of 800~MeV/c$^{2}$ to 1100~MeV/c$^{2}$ chosen in this analysis. The horizontal bands come from the high energy halo muons produced by the secondary beam.} \label{fig:beamline_reco_comp}
\end{figure}

All beamline-triggered events are reconstructed to obtain their individual TOF and momentum as they traverse the beamline monitoring systems. It is possible to distinguish different particle types by utilizing the reconstructed TOF and momentum. This is shown in Fig.~\ref{fig:beamline_reco_comp} with overlapping expectations for particles present in the beam, such as deuterons, protons, kaons, pions, muons, and electrons. Horizontal bands present at TOF intervals of 20~ns are caused by multiple high-energy halo muons produced in the secondary beam far upstream of the tertiary beamline. Their time separation is defined by the primary proton beam structure that strike the upstream and downstream TOF scintillator paddles and create a false TOF measurement for particular beam-triggered events.

\begin{figure}[tb]
	\centering
	\includegraphics[width=0.93\columnwidth]{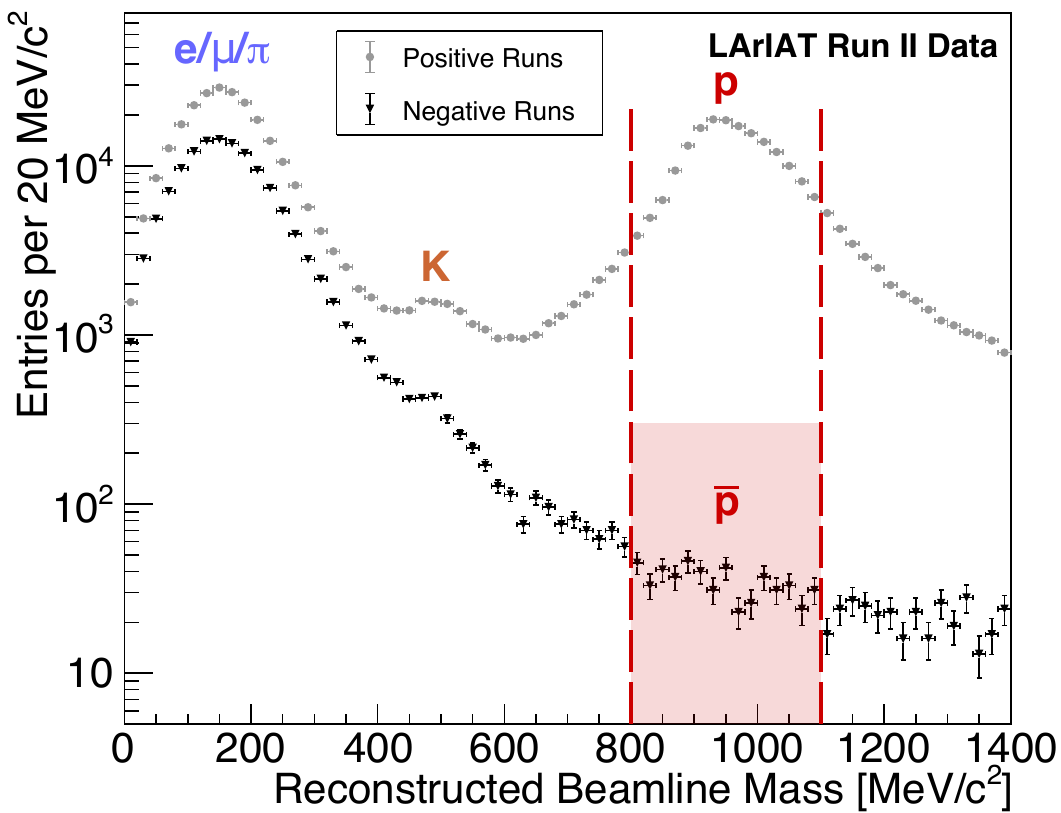}
	\caption{Distribution of the reconstructed beamline mass for positive and negative polarity data runs. The distributions are divided into three major regions in order to distinguish muon/pion-like, kaon-like, and proton-like particle masses. The shaded red region range shows selected antiprotons in the negative polarity data using the positive polarity proton selection cuts.} \label{fig:beamline_reco_mass}
\end{figure}

The invariant beamline mass, $m_b$, for each triggered particle is calculated using its beamline-measured momentum, $p_b$, together with the measured TOF, $t_\text{TOF}$:

\begin{equation}
m_\text{b} = \frac{p_\text{b}}{c}\sqrt{\left( \frac{t_\text{TOF}\cdot c}{l}\right)^2 -1}\label{eq:beamline_mass},
\end{equation}
where $c$ is the speed of light and $l$ is the length traveled by the particle between the two TOF paddles, 6.65~m. The distribution of the reconstructed mass of beamline particles for Run~II is shown in Fig.~\ref{fig:beamline_reco_mass}. Particle types are classified according to three distinct regions as shown in the figure:

\begin{enumerate}
	\item \textbf{electron, muon, pion}:  0~MeV/c$^2$ to 375~MeV/c$^2$,
	\item \textbf{kaon}: 375~MeV/c$^2$ to 650~MeV/c$^2$,
	\item \textbf{proton}: 800~MeV/c$^2$ to 1100~MeV/c$^2$.
\end{enumerate}

The beamline mass selection cuts used to identify antiprotons with their reconstructed momentum and TOF are tuned using the Run~II positive polarity data where there is a large sample of protons ($\sim$100,000). Antiprotons are selected with $m_b$ between 800~MeV/c$^{2}$ and 1100~MeV/c$^{2}$, inclusively, is chosen to include all of the proton-like mass peak while minimizing contamination from kaons at lower mass and deuterons at higher mass. This cut on $m_b$ isolates events in the (anti)proton-like TOF region. The selection is applied to all the data in Run~II taken with \mbox{-100}~A magnet current. The impact of this selection cut is shown in Fig.~\ref{fig:beamline_reco_comp} with the dashed curves outlining the antiproton selection region. Out of the 189,273 events processed in the Run~II negative polarity dataset, 520 are selected as having a mass consistent with an antiproton. The range of selected masses is illustrated by the shaded region in Fig.~\ref{fig:beamline_reco_mass}.

Kaons could mimic antiproton beamline candidates if they have a short track length due to their similar energy deposition profile. To assess potential kaon contamination in the selected events, a combined fit was applied to the negative polarity data from Fig.~\ref{fig:beamline_reco_mass}, with a Gaussian function centered at the fitted kaon mass of 495.2~MeV/c$^{2}$ and on top of a Landau function to model the non-kaon beam related backgrounds in the calculated mass distribution. The backgrounds fit by the Landau are from the high-energy halo muons striking the TOF paddles. Since the events in these horizontal bands are due to random firings of the trigger from muons produced upstream of the tertiary beamline, they feature a high number of tracks in the TPC with no clear single and unobstructed stopping beam-matched track. These are unlikely to contaminate the sample and are removed during the LArTPC selection stage. Based on an extrapolation of the fit into the proton-mass selection window, the expected contamination from kaons is negligible, with $< 10^{-3}$ events expected for the full Run~II \mbox{-100}~A sample.

\section{LArTPC Reconstruction and At-Rest Annihilation Selection}

Following the antiproton beamline reconstruction and selection of Sec.~\ref{sec:beamline}, the second reconstruction and selection phase happens in the LArTPC. In this phase, we use a hybrid manual/semi-automated reconstruction process, culminating in the determination of final candidates on an event-by-event basis using hand-scanning and hand-clustering methods. A ranged-based particle identification (PID) algorithm on both the incoming antiprotons and the outgoing particles from the annihilation vertex.

\subsection{LArTPC Reconstruction \& Simulation}\label{sec:beamline_reco}

LArTPC reconstruction is performed using the LArSoft (version 06.61.00) event processing framework~\cite{larsoft, Acciarri:2019wgd,LArIAT:2021yix} that outputs reconstructed particle objects and their associated kinematics. Particles passing through the LArTPC are reconstructed from ionization signals detected on the wires. These signals are waveforms of ADCs vs.~time ticks. Hits within the waveforms are determined by Gaussian fits to over-threshold pulses found in filtered and deconvolved wire waveforms. Hits are typically clustered together by an automated reconstruction algorithm which reconstructs 3D tracks by time-matching clusters of hits across the two 2D wire-planes~\cite{Baller_2017}. LArIAT's automated reconstruction algorithms were optimized for forward-going particles, and do not reliably reconstruct the more complex, isotropic, multi-track $\bar{p}N$ final-states. Unlike other LArTPCs which typically employ three independent wire readout planes, LArIAT has only two planes, which reduces its ability to use the most up-to-date LArTPC reconstruction algorithms. Therefore, to improve its performance, a manual component to the reconstruction is introduced in this analysis.

\begin{figure}[tb]
		\includegraphics[width=0.95\columnwidth]{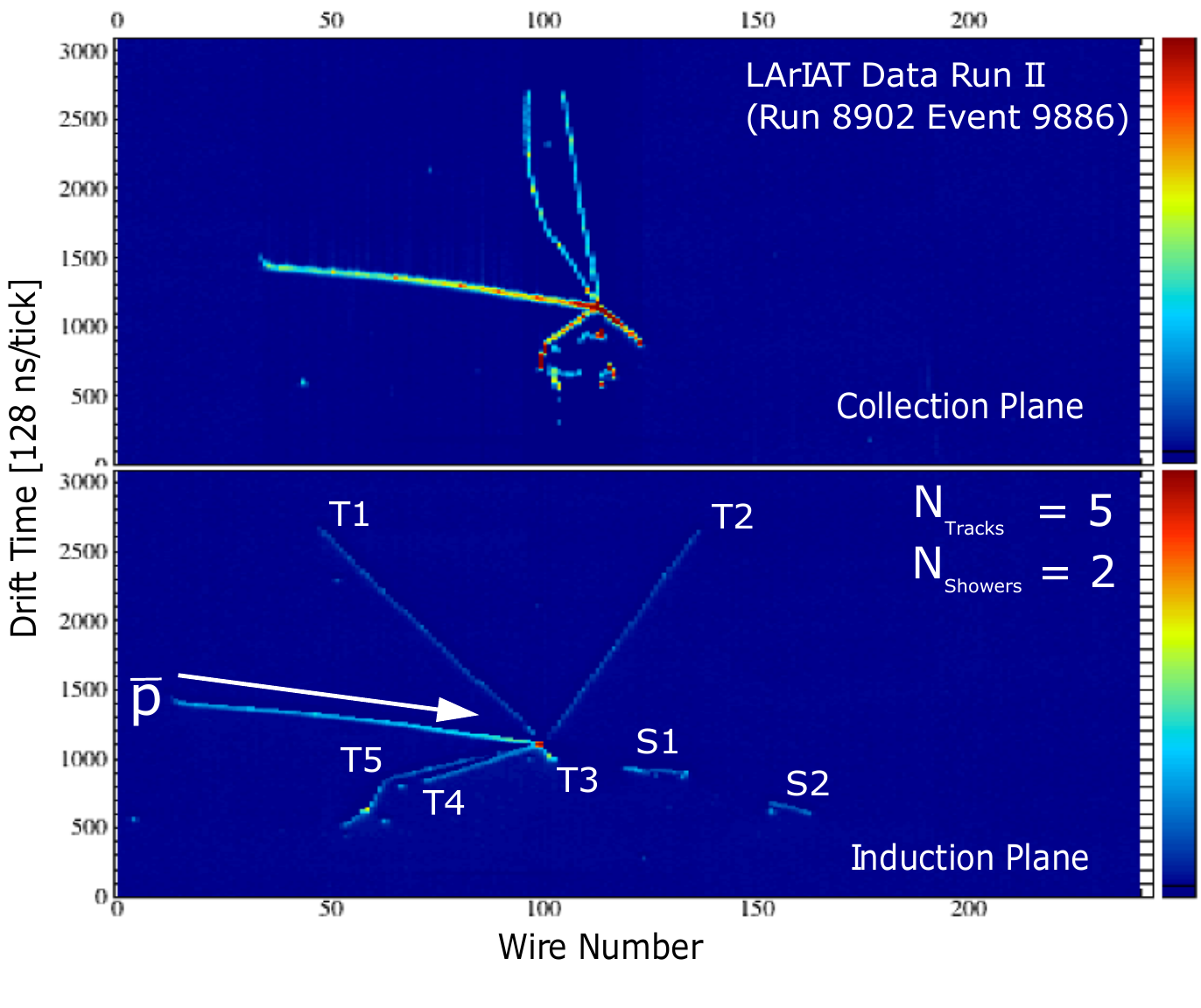}
		\includegraphics[width=0.95\columnwidth]{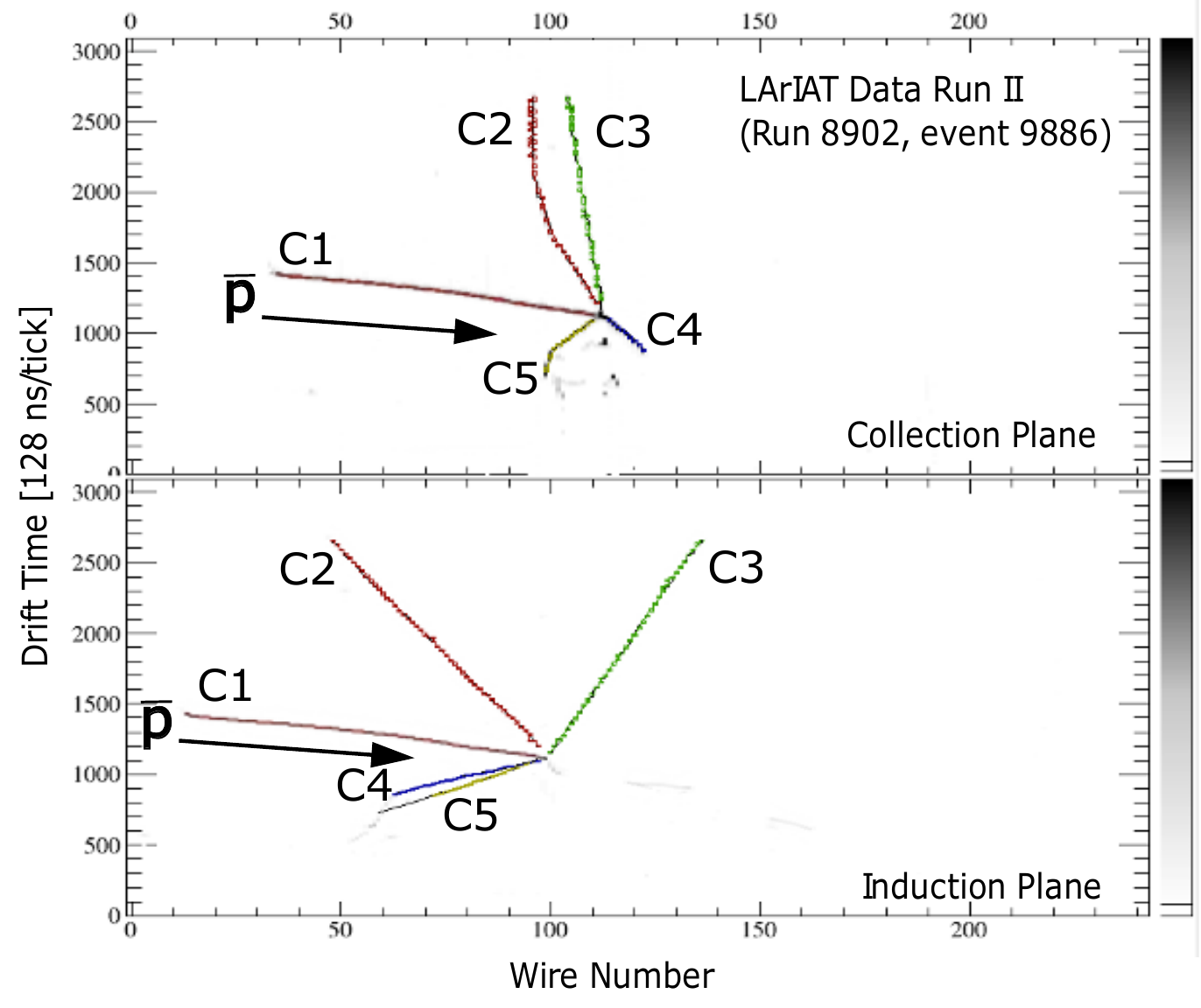}
	\caption{Example of a data antiproton annihilation at rest event (top) and the hand-clustering method applied to the particle track candidates on the same data event (bottom). Hits are manually clustered together for each of the track candidates individually. Each of the different clustered objects, noted from C1 to C5, are passed to the tracking algorithm to form track objects. The incoming antiproton is noted by C1 and the outgoing particles noted from C2 to C5.}
	\label{fig:handcluster_example}
\end{figure}

Contiguous hits that make up the same topological object are manually selected by grouping them together into 2D clusters on each wire plane in a customized event display graphical user interface. These clusters are then used to reconstruct 3D tracks using the standard projection-matching track fitting algorithm in LArSoft~\cite{Yang_pma, Antonello:2012hu}. Hits located directly at the annihilation vertex are excluded from clusters associated with the incoming and outgoing particles to avoid complications in the tracking algorithm. This ensures that hits from the stopping antiproton’s Bragg peak, vertex, or from overlapping outgoing ionization trails are not mistakenly included in the wrong clusters. For candidate incoming antiproton tracks, hand-clustering is treated more inclusively by extending the clustering process as close to the annihilation vertex as possible within two reconstructed hits. An example of the results of individual and manual clustering is shown in Fig.~\ref{fig:handcluster_example} where the different clustered objects are individually grouped in different colors and noted from C1 through C5.

The antiproton is visually identified by looking for tracks which start at the front face of the LArTPC and end within the fiducial volume with a clear increasing d$E$/d$x$ due to the Bragg peak near the endpoint. The reconstructed position at the most downstream point (highest $z$) of the antiproton track is assigned as the annihilation vertex and sets its direction. Outgoing tracks are defined such that their orientation is pointing away from the vertex to ensure effective range-based PID.

In order to gauge the performance of the reconstruction methods, a sample of 1,200 antiprotons entering the LArTPC is simulated. They are generated at the final beamline WC approximately 100~cm upstream of the TPC, with a momentum distribution closely matching what is observed in data by utilizing a Gaussian with similar mean and width. The MC simulation is performed this way to increase the number of $\bar{p}$ annihilations at rest in the active area of the LArTPC, mimicking reality as much as possible. The generated particles decay, interact, and deposit their energy following Geant4~\cite{AGOSTINELLI2003250} (version 10.3.1) using the QGSP\_BERT physics library. To simulate antiproton annihilation on individual nucleons in argon, the Fritiof (FTF)~\cite{ANDERSSON1987289,NILSSONALMQVIST1987387} model is used in this paper.

\subsection{LArTPC Particle Identification and Event Selection}\label{sec:lartpc_selection}

Selected beamline events are classified using two hand-scanned selection criteria in the LArTPC:
\begin{enumerate}
    \item A clear, non-overlapping, incoming beamline particle track must come to a stop in the active volume, producing a Bragg peak characterized by a visible increase in charge density along the track (indicated by color in the event display).
    \item Outgoing particles must start at the end of the incoming beamline particle and travel away from the annihilation vertex. These particles are not required to be contained.
\end{enumerate}

The first criterion removes events where the annihilation might have occurred before entering the active region of the LArTPC resulting, in some cases, in an empty event. In data, this criterion additionally removes events containing significant pile-up muon activity which may produce ionization that overlaps and interferes with the actual triggered antiproton track. These pile-ups are caused by additional tracks from muons produced in the upstream beamline elements traveling alongside the true beamline candidates. Since annihilation at rest produces outgoing particles isotropically, which are required to be observed from the second criterion, it is possible that tracks in the LArTPC overlap with the incoming stopping antiproton. The Bragg peak of the selected events ensures that the particle is coming to rest and its track information is passed to the d$E$/d$x$-based PID algorithm to identify the particle type.

The particle associated with a reconstructed track inside the LArTPC is identified by examining the calorimetry information for the track’s reconstructed hits. This is done by looking at the amount of energy deposited per unit length (d$E$/d$x$) along the 3D track object and comparing it to the expected values of d$E$/d$x$ for different particle types. In this analysis, particles can be grouped as: proton-like, kaon-like, pion/muon-like, and minimum-ionizing particles without a Bragg peak (non-stopping-like). The PID method consists of calculating a $\chi^2/\text{NDF}$ between a track’s modified residual range profile and expected d$E$/d$x$ profiles for different particle types. The number of degrees of freedom (NDF) is the number of reconstructed hits used to calculate the total $\chi^2$.

To improve the performance of the algorithm, which relies on a well-defined and clear Bragg peak, reconstructed hits at or near the annihilation vertex are omitted, as these can bias the $\chi^2$ calculation. Specifically, hits within 1~cm of the vertex are excluded as the end of the antiproton Bragg peak can start to merge with the annihilation vertex and overlap with the ionization from outgoing particles. The PID method accounts for this potential under-clustering of the track, which can lead to shorter reconstructed lengths. To mitigate this effect, a correction is applied to the residual range, adjusting it by up to two reconstructed hits—equivalent to 1-2~cm due to the wire-based readout and discretization effects.

In order to determine the selection efficiency, the criteria definition for an ``at-rest’’ annihilation signal at truth-level is defined as an antiproton that annihilates while retaining an amount of kinetic energy less than or comparable to that carried by any one of its constituent quarks in the antiproton’s rest frame, equivalent to $\sim$100~MeV due to internal Fermi motion. Since the energy imparted to the final-state annihilation products is the sum of the antiproton’s rest mass energy, its kinetic energy at the moment of interaction, and this random quark motion, this requirement on the true signal ensures events whose final-state metrics (energy and track multiplicity) are comparable to true at-rest annihilation within $\sim$5-10\%.

After visually scanning 1,200 simulated antiproton events, 80 stopping candidates were selected. Of these candidates selected through hand-scanning, 69 truly annihilate at rest, 9 retain some residual kinetic energy less than 60~MeV, while the remaining 2 annihilate with nearly 200~MeV of remaining energy. Based on the signal definition above, the mis-ID rate for visual inspection is therefore estimated to be $2.5\pm1.8$\%. Of the initial 1,200 events, 140 are considered signal as they have true final kinetic energy of less than 100~MeV. This selection of antiproton annihilation at rest events results in a selection efficiency of 56\% and a purity of 98\%.

\begin{figure}[tb]
	\centering
    \includegraphics[width=\columnwidth]{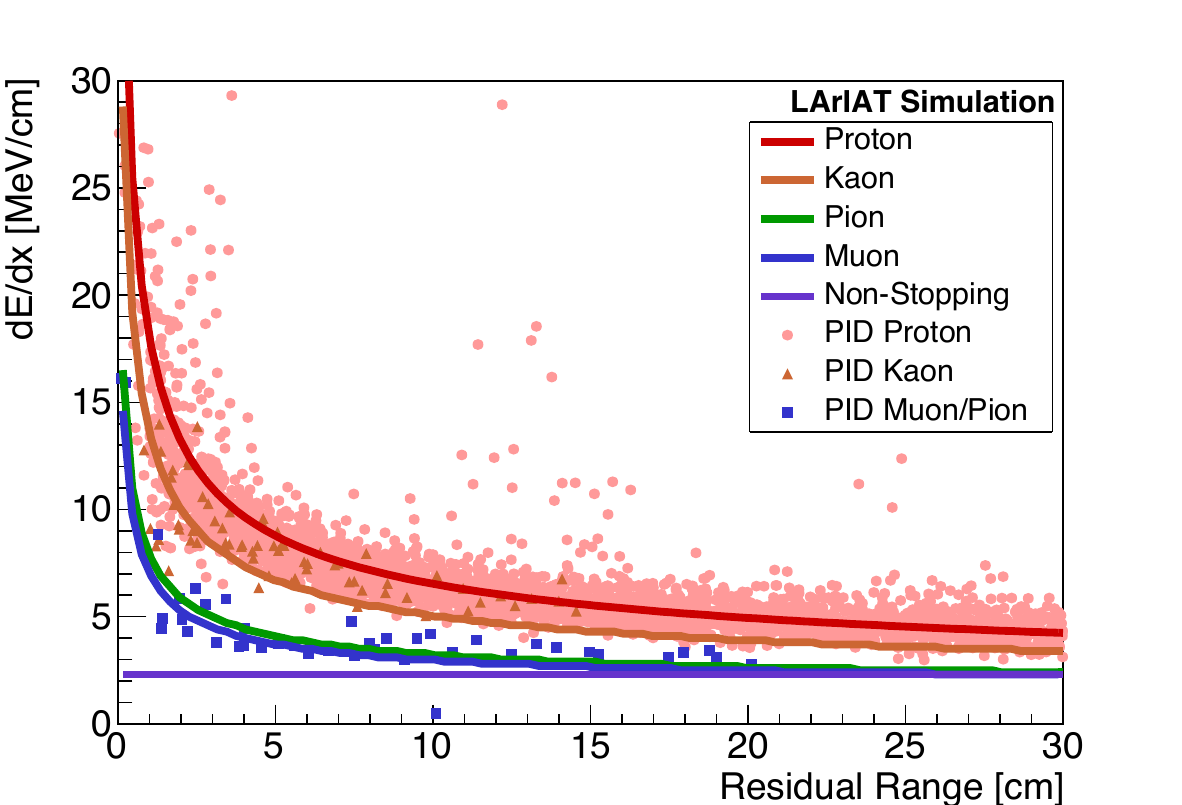}
    \includegraphics[width=\columnwidth]{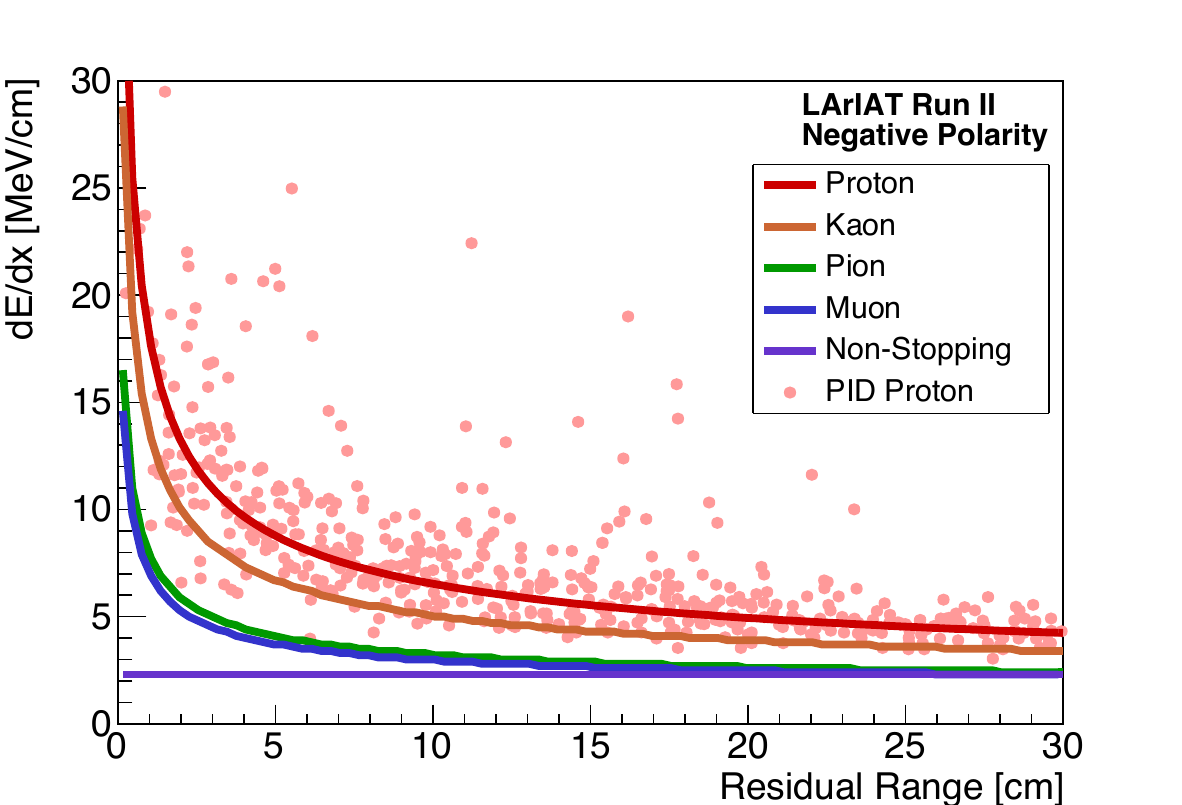}
    \caption{Distributions of the measured d$E$/d$x$ for points along candidate tracks as a function of the residual range for all 80 simulation candidates passing the selection cuts (top), as well as for all 11 antiproton candidates from data (bottom).} \label{fig:dedx_rr_mc_data}
\end{figure}

All of the 80 MC signal candidates are also processed through the PID algorithm. The results indicate that 70 of the 80 events are identified as proton-like, 6 as kaon-like, and 4 as stopping muon/pion-like. To evaluate the performance of the PID algorithm, the d$E$/d$x$ distributions for the three categories of antiproton candidates are compared with the expected values, as shown in top of Fig.~\ref{fig:dedx_rr_mc_data}. The proton-like candidates closely align with the proton expectation curve. In comparison, for proton tracks that interact before coming to a full stop, or have a mis-reconstructed endpoint that is placed upstream of the true endpoint, the d$E$/d$x$ profile tends to more closely resemble a kaon-like hypothesis. The stopping muon/pion-like events, which correspond well with the pion/muon expectation curve, are rejected from our sample. The two antiproton candidates that have over 200 MeV of kinetic energy were identified as 1 kaon-like and 1 muon/pion-like and would be removed at this stage. Based on these observations, all 10 non-proton-like candidates are excluded from the sample, resulting in a PID efficiency of 88\%. 

\begin{figure}[tb]
	\centering
	\includegraphics[width=0.95\columnwidth]{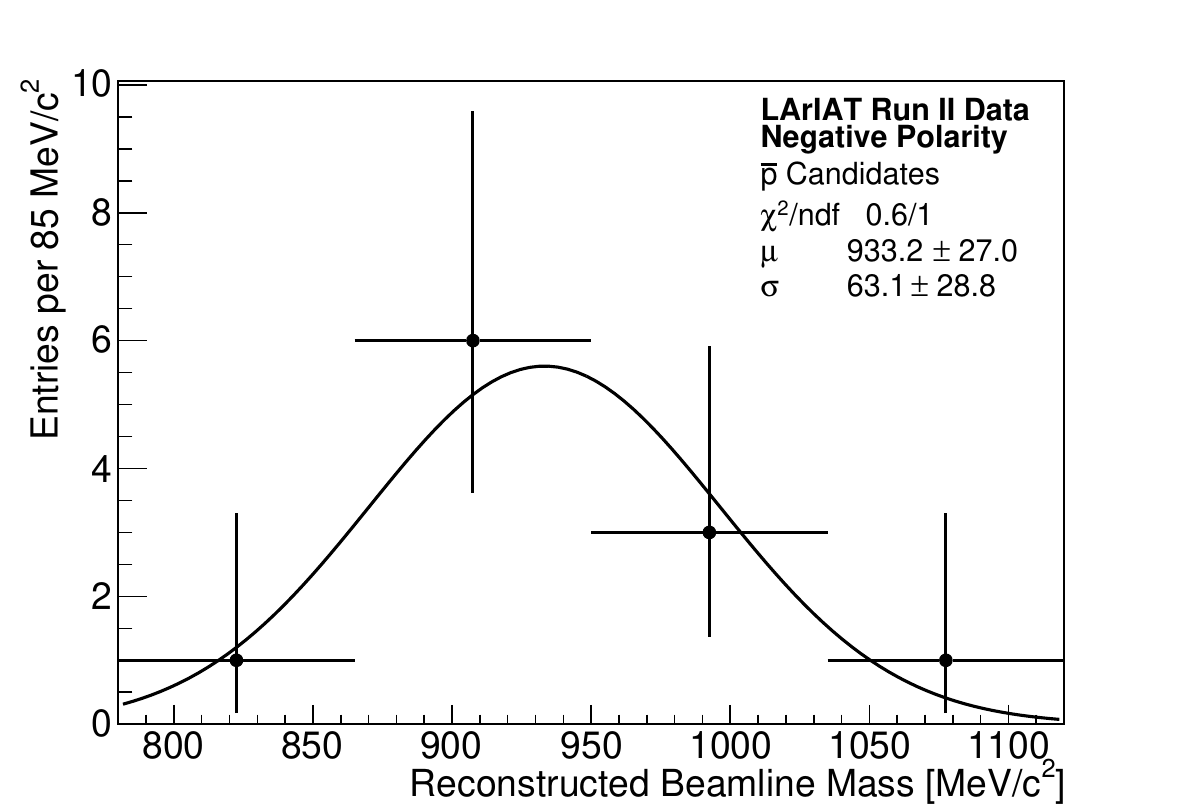}
	\caption{Distribution of the reconstructed beamline mass of the 11 stopping antiproton candidates with their statistical uncertainty with a fitted Gaussian overlaid resulting in a mass peak of 933.2 $\pm$ 27.0~MeV/c$^2$.} \label{fig:candidates_beamlinemass}
\end{figure}

Following the established selection criteria, the number of data events from the data period described in Sec.~\ref{sec:lariat_detector} is reduced from 520 initial candidates in the selected mass range, down to 11 candidate stopping antiprotons annihilating at rest. The reconstructed beamline mass distribution shown in Fig.~\ref{fig:candidates_beamlinemass} with a fitted Gaussian centered at 933.2 $\pm$ 27.0~MeV/c$^2$ suggests that these are all true antiprotons. All 11 candidates are also identified as proton-like using the PID $\chi^2$ calculation. The d$E$/d$x$ as a function of the residual range distribution of the antiproton candidates is shown in bottom of Fig.~\ref{fig:dedx_rr_mc_data}. All 11 candidates align well with the proton-like expected curve confirming that they are expected to be true antiproton annihilations at rest given the small misidentification rate of $2.5\pm1.8$\% evaluated from the simulation.

A summary of the effect of each selection cut is shown in Table \ref{table:selection_antiproton}. High energy muon pile-ups are not simulated, so a difference in the relative number of events removed from data and simulation is expected for the second cut. A collection of selected antiproton annihilation event displays from data and simulation is provided in Appendix \ref{appendix:pbar_events}.

\begin{table}[tb]
    \renewcommand{\arraystretch}{1.5}
    \centering
 \setlength{\tabcolsep}{12pt}
    \begin{tabular}{l c c}
   	 \hline\hline
            \toprule
   	 Selection Cut & Data & $\bar{p}$ Simulation \\ 
   	 \hline
   	 mass candidates
    	& \makecell{520}
    	& \makecell{1,200}
    	\\
    	\makecell[l]{pile-up or empty events}
    	& \makecell{145}
    	& \makecell{800}
    	\\
                \makecell[l]{stopping candidates}
    	& \makecell{11}
    	& \makecell{80}
    	\\
         \makecell[l]{PID}
    	& \makecell{11}
    	& \makecell{70}
    	\\
   	 \hline
    \end{tabular}
    \caption{Summary of the selection cuts for the antiproton annihilation at rest for both the beamline and LArTPC stages with the first line corresponding to the beamline selection for data, and all of the simulated events. The second and third line corresponds to the hand-scanned selection while the fourth and final line to the PID using the LArTPC reconstructed hits.}
    \label{table:selection_antiproton}
\end{table}

\section{Final-State Reconstruction Results}

\begin{figure}[tb]
	\centering
	\includegraphics[width=\columnwidth]{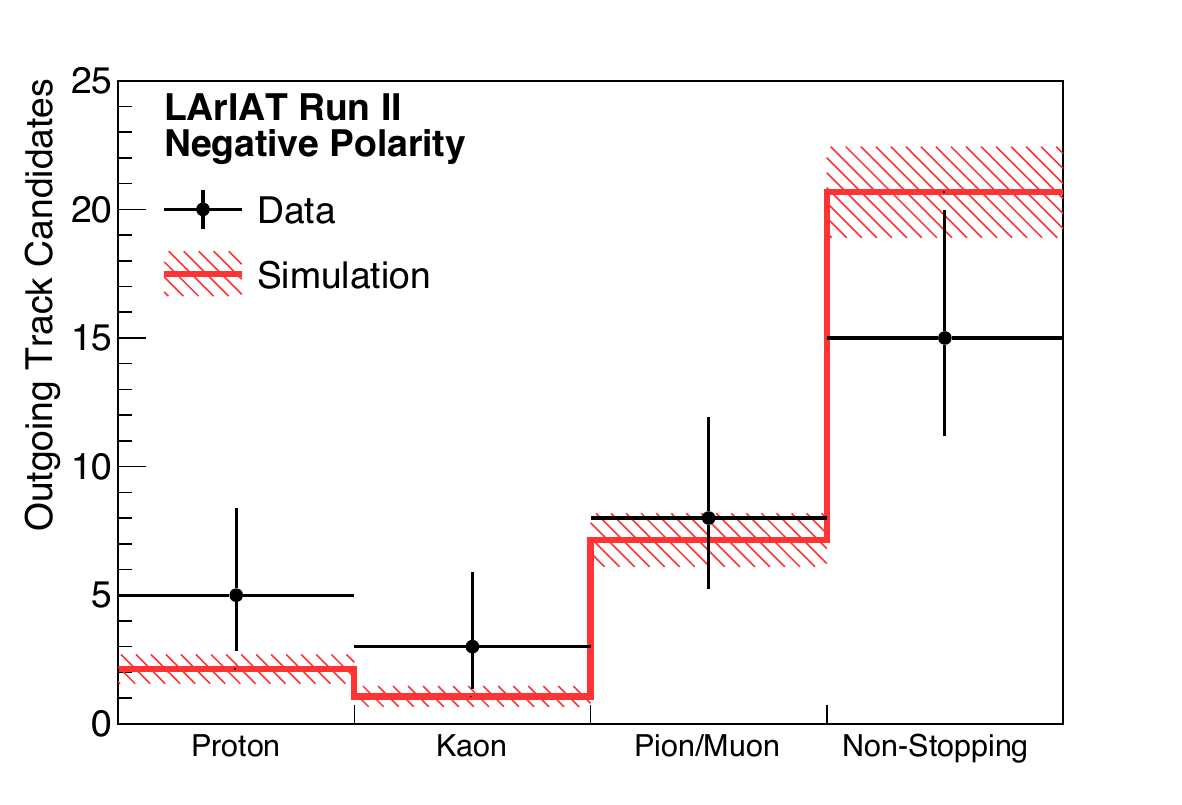}
	\caption{Distribution of the outgoing particle type from the PID algorithm. The number of signal antiproton tracks from the simulation was normalized to the data, and both are shown with their respective statistical uncertainty.} \label{fig:candidates_out_chi2_data_mc}
\end{figure}

As with the incoming antiproton candidates, the particle tracks leaving of the annihilation vertex are passed through the PID algorithm to determine their particle type with the results for the data and simulated events shown in Fig.~\ref{fig:candidates_out_chi2_data_mc}. The simulation was normalized to the number of data antiproton events to allow direct comparisons. The data and simulation are consistent within statistical uncertainty. Half of the outgoing particles have an end point within 1~cm from an edge in which case they are considered to be exiting the LArTPC. This makes it impossible to perform a definitive PID based on the d$E$/d$x$ of these tracks and these contribute to the last bin in Fig.~\ref{fig:candidates_out_chi2_data_mc}.

The final-state outgoing track multiplicity is obtained in two different ways: by hand-counting the number of tracks leaving the annihilation vertex, and by counting the number of reconstructed tracks determined by the semi-automated algorithm. This allows for a comparison between the two methods and is possible because of the small size of the final candidate sample in this analysis. Due to complications from manually clustering multiple tracks emanating from a common vertex, hits near the vertex were often excluded, as explained in Sec.~\ref{sec:lartpc_selection}. Therefore, a generous maximum track-to-vertex distance of 2~cm was allowed for outgoing track candidates, with a required span of five wires to ensure that tracks extend far enough from the vertex region.

\begin{figure}[tb]
	\centering
	\includegraphics[width=\columnwidth]{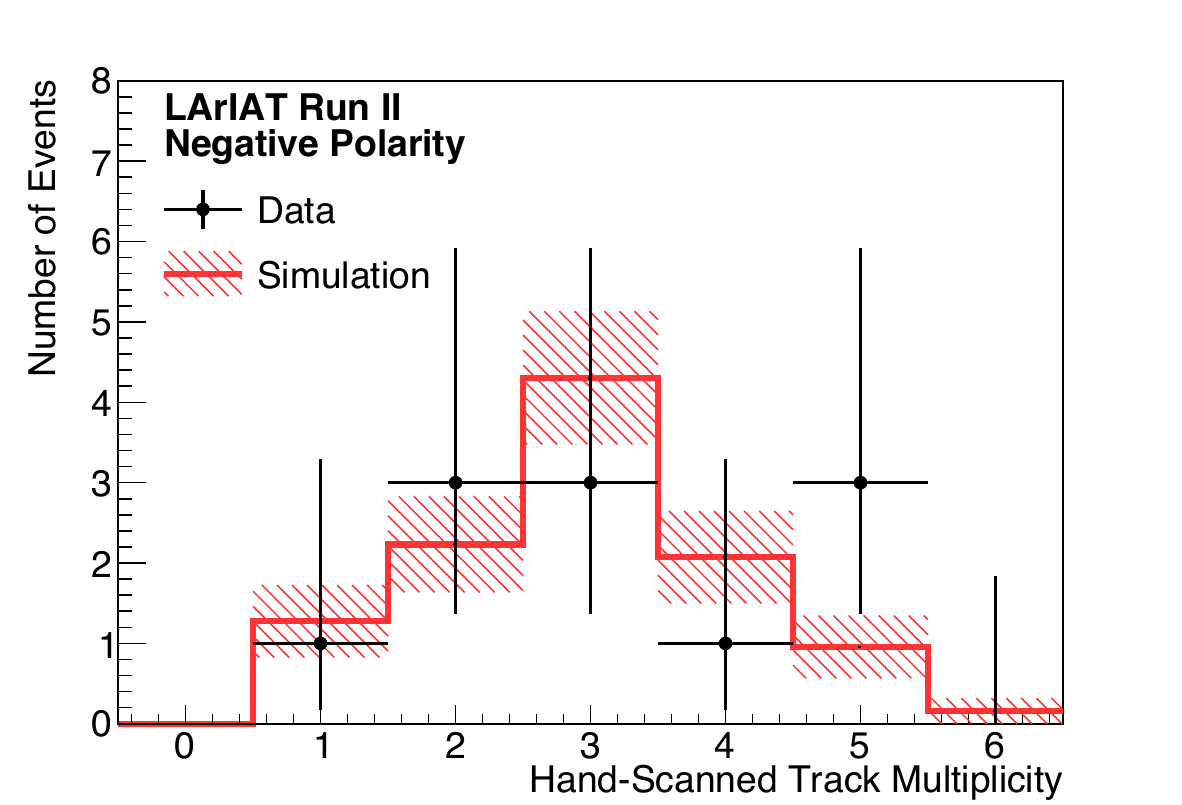}
        \includegraphics[width=\columnwidth]{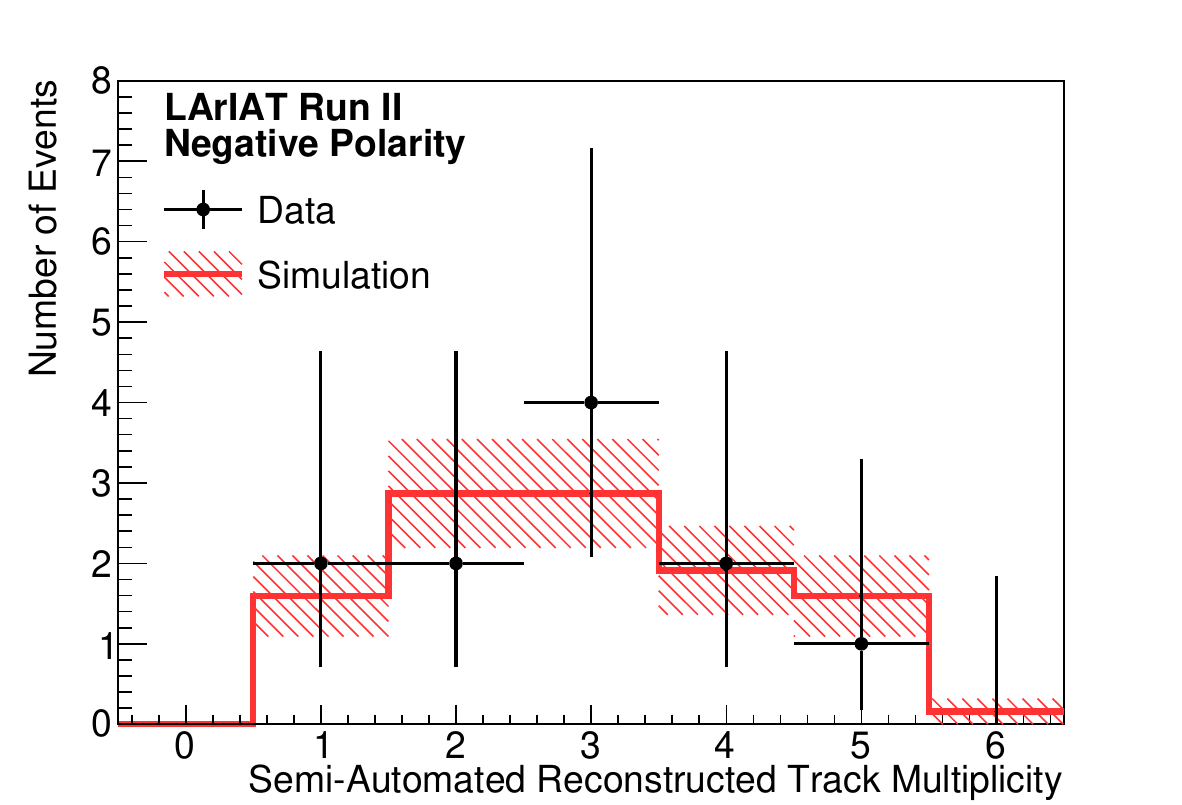}
	\caption{Distribution of charged particle track multiplicity for outgoing particles from antiproton annihilation at rest: (top) the hand-scanning method, and (bottom) semi-automated reconstruction algorithm method. Inefficiencies in the tracking part of the reconstruction reduce the overall number of reconstructed tracks. The simulated sample is normalized to the number of events in data.} \label{fig:multiplicity_final}
\end{figure}

The track multiplicity distributions comparing simulation to data for both the hand-scanning and semi-automated reconstruction methods are shown in Fig.~\ref{fig:multiplicity_final}. The data distributions are consistent with the simulation within statistical uncertainty.

A summary of the calculated means, the uncertainty of the means and the root mean squares (RMS) values for both track multiplicities for the hand-scanning and semi-automated reconstruction is shown in Table \ref{table:track_summary}. The mean for the semi-automated reconstructed counting method is slightly lower than for the hand-scanned due to reconstruction failures and inefficiencies. All calculated means agree within uncertainties.

\begin{table}[tb]
    \renewcommand{\arraystretch}{1.5}
    \centering
 \setlength{\tabcolsep}{12pt}
    \begin{tabular}{l c c}
   	 \hline\hline
   	 Method & Data & Simulation \\ [0.1ex]
   	 \hline \\[-1.0em]
   	 Hand-scanning
    	& \makecell{$3.2\pm0.4$\\$\sigma=1.3$}
    	& \makecell{$3.01\pm0.14$\\$\sigma=1.15$}
    	\\ [1em]
    	\makecell[l]{Semi-automated\\reconstruction}
    	& \makecell{$2.8\pm0.4$\\$\sigma=1.2$}
    	& \makecell{$2.96\pm0.16$\\$\sigma=1.31$}
    	\\[1.0em]
   	 \hline
    \end{tabular}
    \caption{A summary of the mean with the uncertainty on the mean and RMS ($\sigma$) of measured final-state track multiplicity for the antiproton annihilation candidate samples in LArIAT.}
    \label{table:track_summary}
\end{table}

\begin{figure}[t!]
	\centering
	\includegraphics[width=\columnwidth]{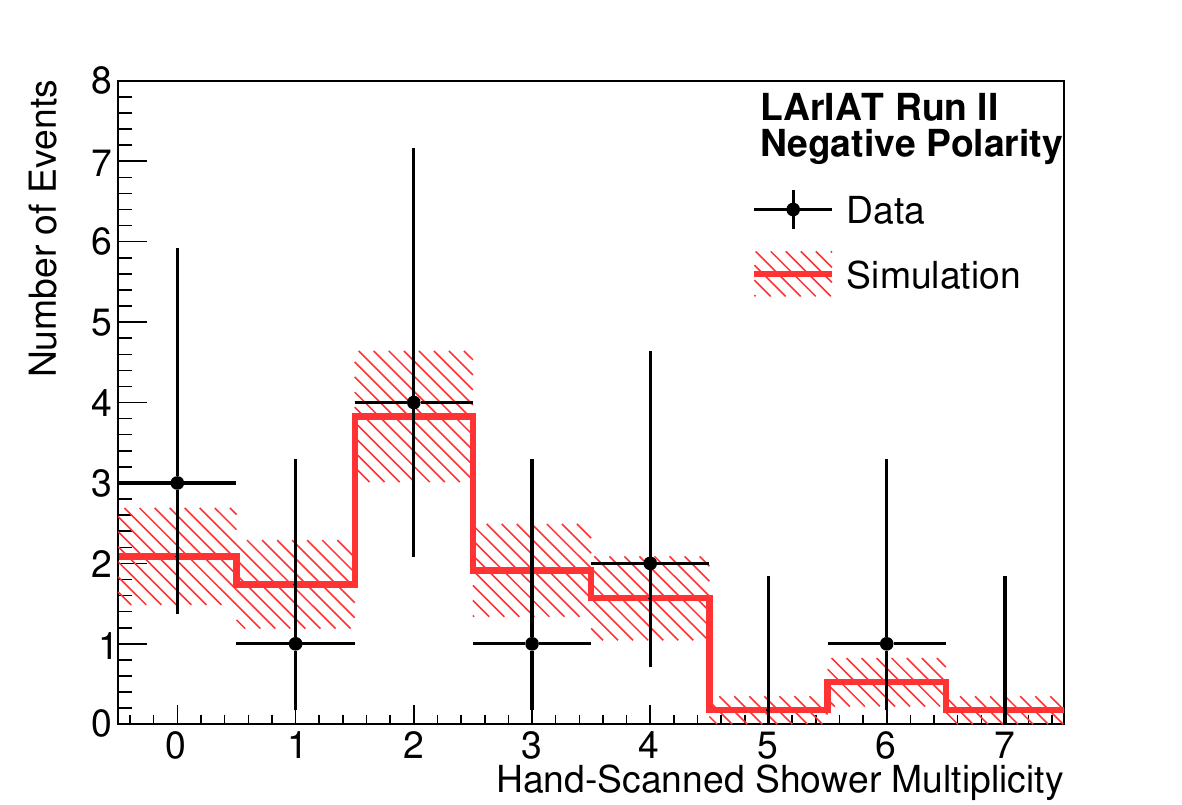}
	\caption{Distribution of shower multiplicity from antiprotons annihilating at rest using the hand-scanning method. The simulated sample is normalized to the number of events in data.} \label{fig:multiplicity_final_showers}
\end{figure}

Antiproton annihilation may also produce $\pi^{0}$ and $\eta^{(')}$~\cite{Amsler:1997up, CrystalBarrel:2003uej}, both of which decay into photon showers ($\gamma$). These particles decay promptly which can result in a pair of showers pointing directly back to the annihilation vertex at some distance. Showers pointing back to the vertex are counted through hand-scanning but the ones pointing back to an outgoing track are not counted as they are considered to originate from one of the outgoing particle tracks rather than directly from the annihilation. Due to the conversion length of the $\gamma$ (greater than 29~cm~\cite{MicroBooNE:2019rgx}), and the small size of LArIAT, $\gamma$ can exit the LArTPC before producing a visible shower. The simulation predicts that $>$40\% of the $\gamma$ exit the LArTPC before interacting. This results in an observed lower limit on the number of showers coming from the annihilations. Similar to the track multiplicity, the mean shower multiplicity for data (simulation) corresponds to $2.2 \pm 0.5$ ($2.2 \pm 0.5$) showers with RMS of 1.8 (1.6). The distributions are shown in Fig.~\ref{fig:multiplicity_final_showers} and data and simulation are consistent within statistical uncertainty.

\section{Conclusions}

LArIAT has observed the first antiproton annihilation at rest interactions on argon using a LArTPC detector in a test beam. A total of 11 candidates of antiproton annihilation at rest are identified in LArIAT's Run~II negative polarity data using various beamline and LArTPC selections. Distributions of final state particle multiplicity and type were compared to MC simulations using Geant4’s FTF model for surface nucleon annihilation. For data, a mean of $3.2\pm0.4$ tracks per annihilation is measured with a RMS of 1.3 for the hand-scanned method and $3.2\pm0.4$ tracks with a RMS of 1.2 for the semi-automated method. Measurements of shower multiplicity are also consistent between data and simulation, with data showing a mean of $2.2\pm0.5$ showers within LArIAT’s active volume with RMS of 1.8. The predictions from the models used in Geant4 demonstrate consistency with the data, as they fall within the statistical uncertainty for the low-statistics sample used in this paper. The methods outlined in this paper offer a foundation for higher-statistics measurements in larger LArTPCs, which could further assess the models' agreement and potentially inform improvements in modeling approaches by adding a new nucleus to the list of observed antiproton annihilation at rest events.

\section{\label{sec:Results}Acknowledgments}
This document was prepared by the LArIAT collaboration using the resources of the Fermi National Accelerator Laboratory (Fermilab), a U.S. Department of Energy, Office of Science, HEP User Facility. Fermilab is managed by Fermi Research Alliance, LLC (FRA), acting under Contract No. DE-AC02-07CH11359. We also gratefully acknowledge the support of the National Science Foundation; Brazil CNPq grant number 233511/2014-8, Coordena\c{c}\~ao de Aperfei\c{c}oamento de Pessoal de N\'ivel Superior - Brazil (CAPES) - Finance Code 001, S\~ao Paulo Research Foundation - FAPESP (BR) grant number 16/22738-0; Polish National Science Centre grant Dec-2013/09/N/ST2/02793; the Science and Technology Facilities Council (STFC), part of the United Kingdom Research and Innovation; The Royal Society (United Kingdom); Marie Sklodowska-Curie grant agreements numbers 822185 and 892933; STFC grant number ST/W003945/1; and the JSPS grant-in-aid (Grant Number 25105008), Japan. The collaboration extends a special thank you to the coordinators and technicians of the Fermilab Test Beam Facility, without whom none of this work would have been possible. The authors want to also thank Dr. Adam Lister for insightful discussion throughout the review of this paper.
\bibliography{references} 
\appendix
\section{Selected Data Antiproton Candidates}\label{appendix:pbar_events}

\begin{figure*}[t]
    \centering

    \subfloat[\label{subfig:a}]{%
        \includegraphics[width=0.9\columnwidth]{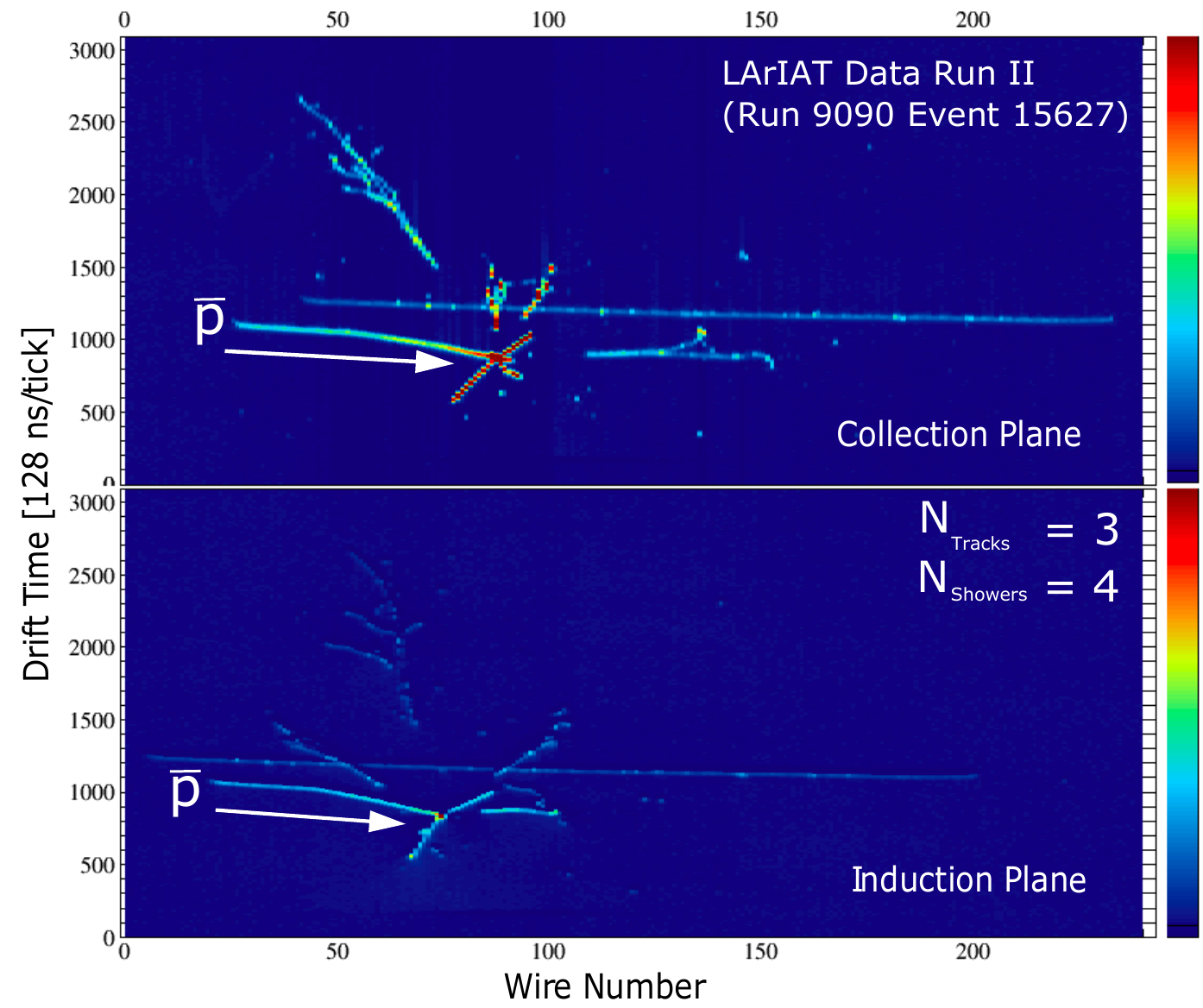}
    }
    \hfill
    \subfloat[\label{subfig:b}]{%
        \includegraphics[width=0.9\columnwidth]{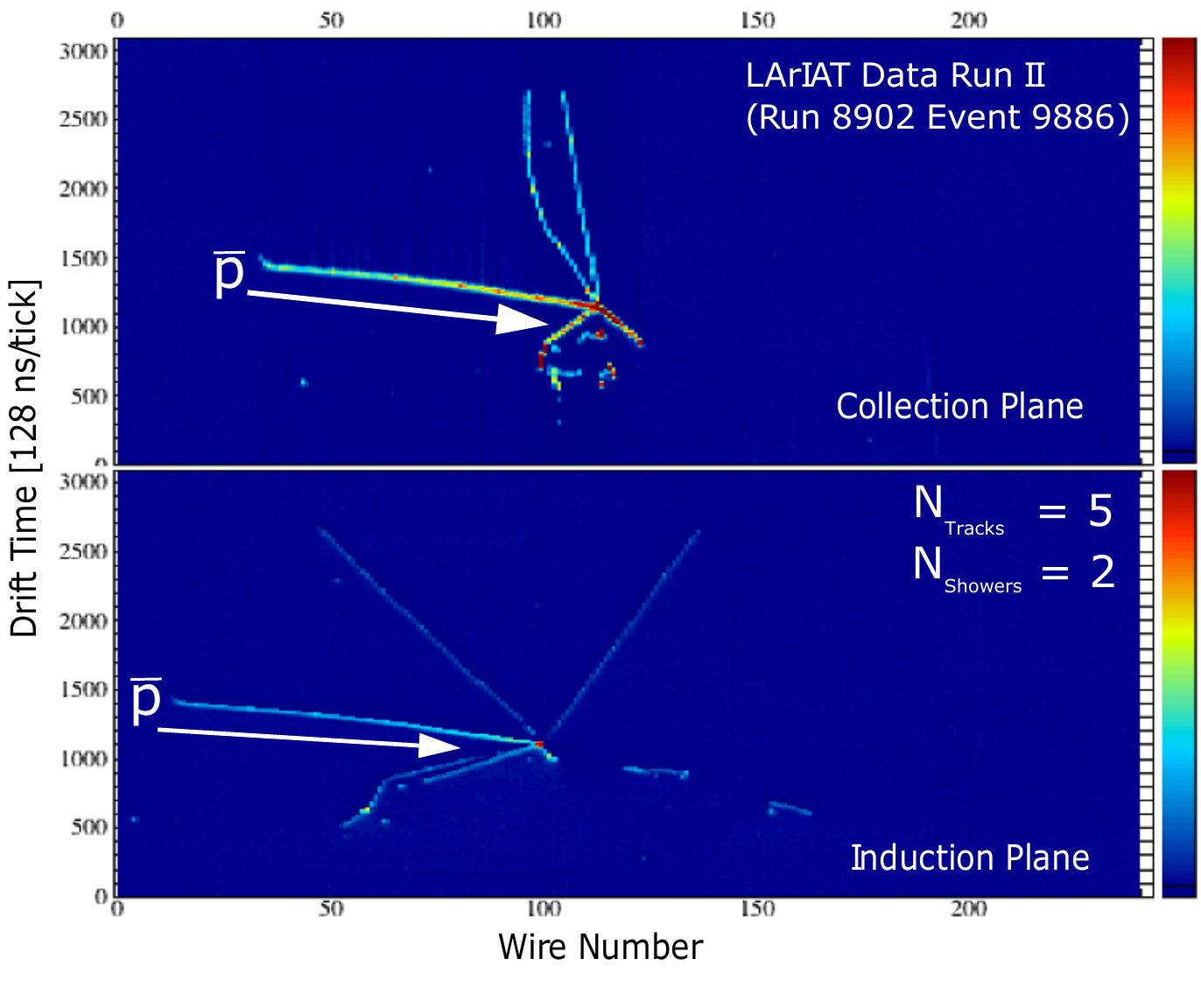}
    }
    \hfill
    \subfloat[\label{subfig:c}]{%
        \includegraphics[width=0.9\columnwidth]{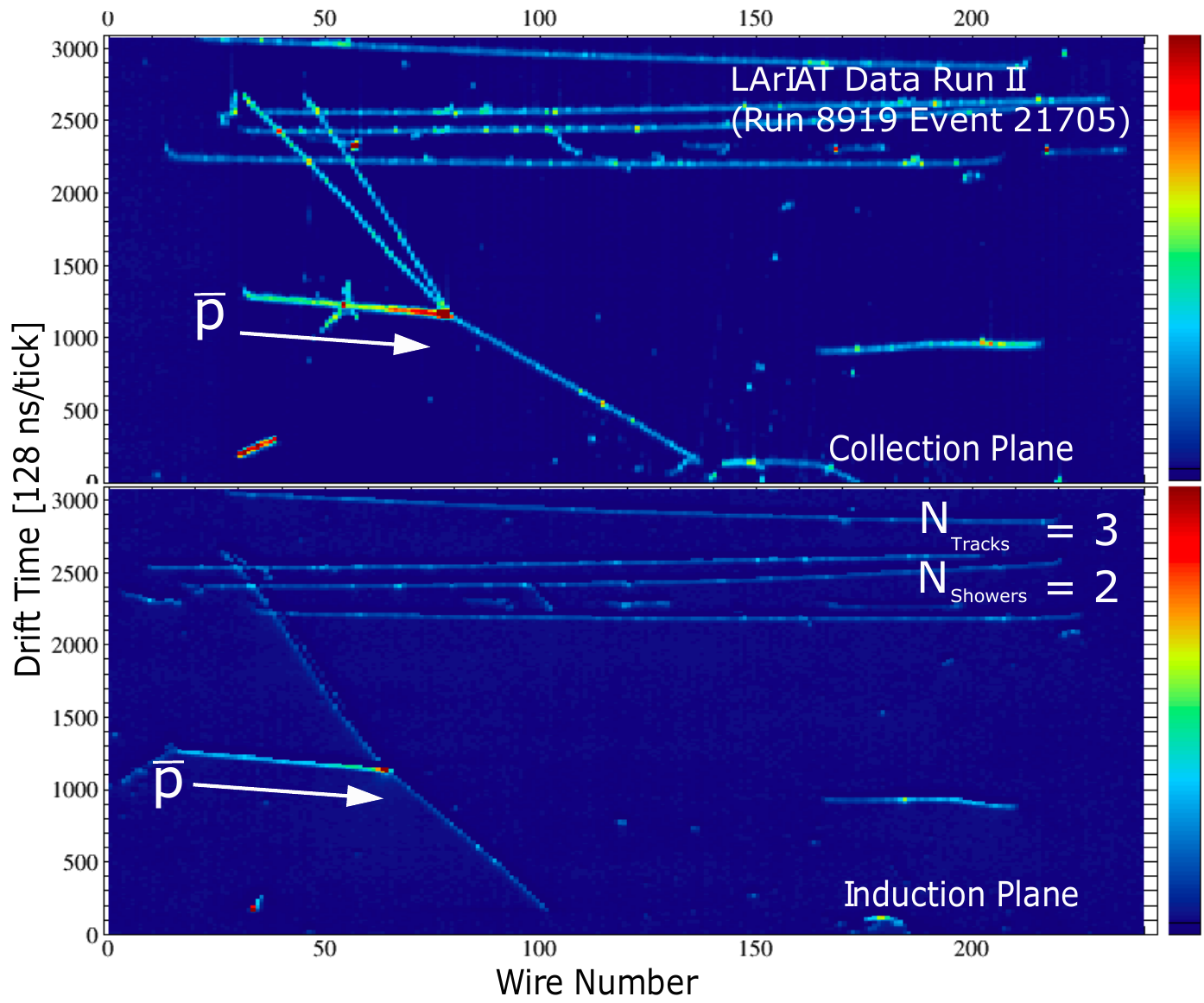}
    }
    \hfill
    \subfloat[\label{subfig:d}]{%
        \includegraphics[width=0.9\columnwidth]{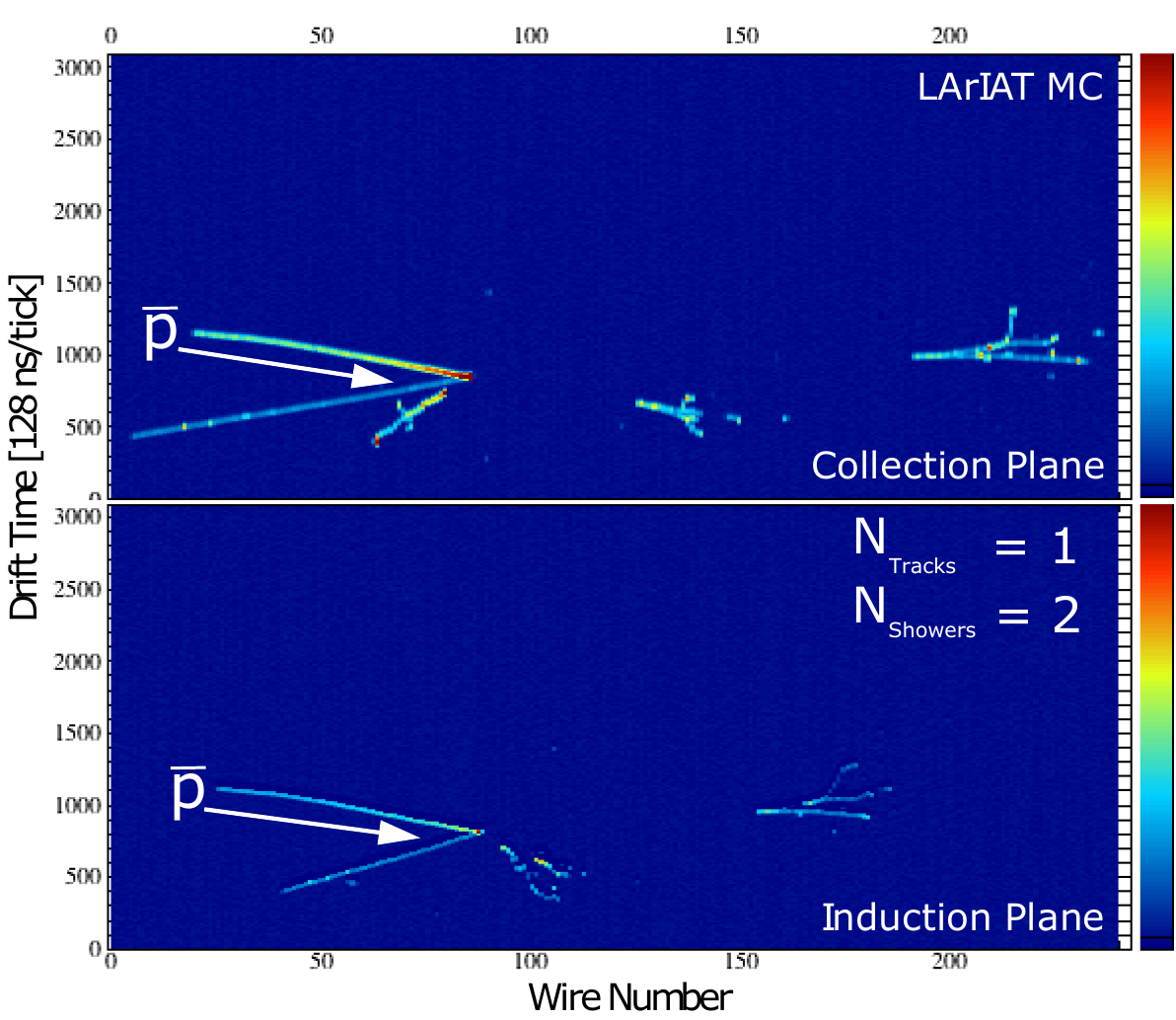}
    }
    \hfill
    \subfloat[\label{subfig:e}]{%
        \includegraphics[width=0.9\columnwidth]{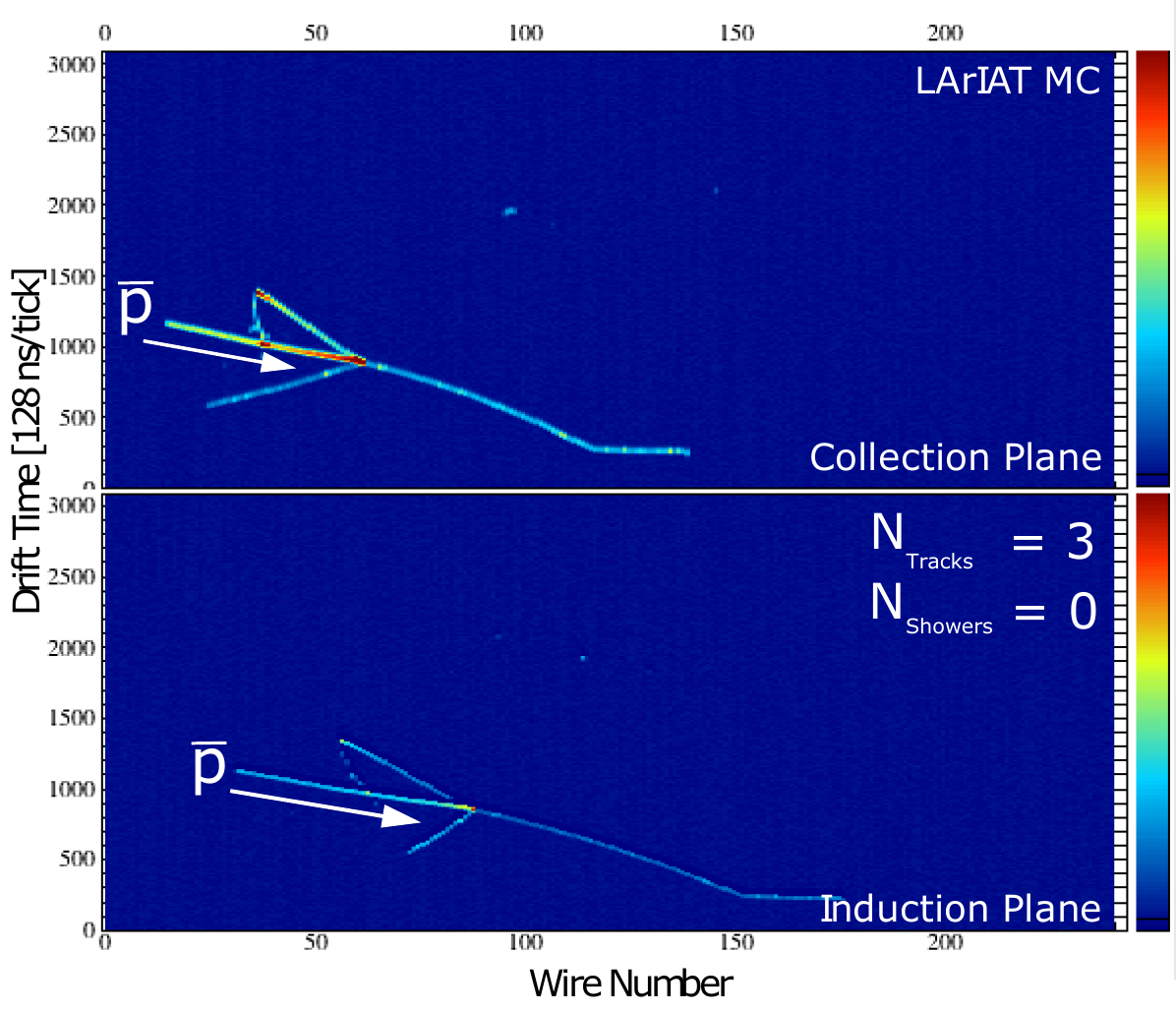}
    }
    \hfill
    \subfloat[\label{subfig:f}]
    {
        \includegraphics[width=0.9\columnwidth]{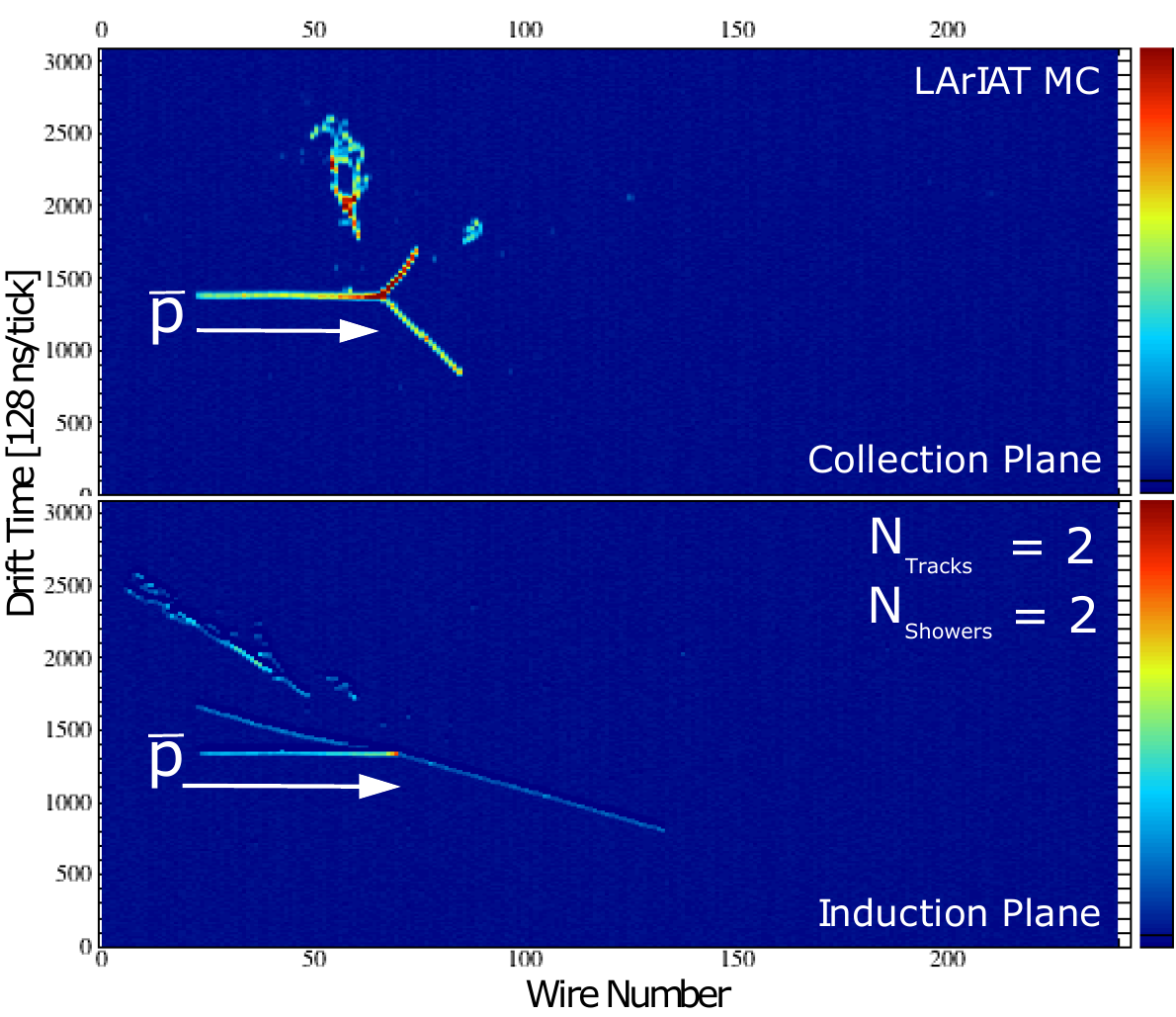}
    }
    \hfill
        
    \caption{Antiproton annihilation data (a)-(c) candidates and MC (d)-(f) obtained from the selection. The incoming identified antiprotons are marked with $\bar{p}$ and their direction with the arrows both in white.}
    \label{fig:all_evds}
\end{figure*}

Figure \ref{fig:all_evds} shows a selection of representative data and MC event displays for antiproton annihilation at rest. The highly ionizing track traveling from the middle-left towards the center of the panels are the incoming antiproton marked in white with $\bar{p}$ and arrow for the direction. At the end of the antiproton's Bragg peak, outgoing tracks are seen exiting directly from the annihilation vertex, while outgoing showers exhibit a gap between the vertex and their starting point but are still oriented directly towards the vertex. The number of hand-scanned tracks (N$_{\mathrm{tracks}}$) and showers (N$_{\mathrm{showers}}$) for these events are indicated in the event displays. In the case of (a) and (c), the relatively straight tracks near the top half of the collection and induction plane panes are high-energy muons produced from the pion decays in the secondary beam far upstream of the LArIAT tertiary beamline, which can contaminate beamline-triggered readouts with multiple extra tracks unassociated with the tertiary beamline.

\end{document}